\documentstyle[prb,aps,multicol,epsf]{revtex}
\begin{document}

\newcommand{\IM}{\mbox{Im}}
\newcommand{\RE}{\mbox{Re}}

\tolerance 10000

\draft

\title{Temperature Dependence of Transport Coefficients in Liquid and
Amorphous Metals}

\author{Ulrich Werner$^{1}$ and Raymond Fr\'esard$^{2,*,+}$}
\vspace*{0.5truecm}

\address{
$^{1}$Institut f{\"u}r Theorie der Kondensierten Materie,
Physikhochhaus, Universit{\"a}t Karlsruhe\\ Postfach 6980, 76128
Karlsruhe, 
Federal Republic of Germany\\
$^2$ Institut de Physique, Universit\'e de Neuch\^atel, A.-L. Breguet 1
CH-2000 Neuch\^atel}

\date{\today}
\maketitle
\widetext

\begin{abstract}
\begin{center}
\parbox{14cm}{
We apply the muffin-tin effective medium approximation to calculate
the temperature dependence of the resistivity and of the thermopower
of amorphous and liquid metals. The results show unambiguously that a
large resistivity  is accompanied by a negative temperature
coefficient, in agreement with the experimental situation. This
behavior is shown to result from a pseudo-gap which opens in the
1-particle spectrum due to strong scattering at the quasi zone
boundary.
In turn the thermopower is found to have non-trivial density and temperature 
dependences.} 
\end{center}
\end{abstract}

\pacs{
\hspace{1.9cm}
PACS number(s): 71.23, 71.55, 72.15 }
\begin{multicols}{2}

\section{Introduction}
\noindent
The influence of a disordered potential on the motion of electrons
remains an interesting problem despite of decades of work. The
aim of solving this problem is  
to elucidate the properties, if not the existence, of a whole range of
materials, such as doped semi-conductors, doped Mott insulators,
amorphous and liquid metals, to quote a few.
The classes of materials are not only distinguished by the state of
their parent compounds, but also by the type of disorder and the
influence of the electron-electron interaction. Even if the latter does
certainly play a key role for the  doped Mott insulators, it is
commonly accepted that its role is much reduced in most amorphous and
liquid metal systems. Types of disorder can be, among others,
``diagonal'' as in conventional modeling of doped semi-conductors,
which can lead to the localization transition \cite{aral} (for a
recent review see \cite{VW}), or topological, or both. The strength of
the disorder is an equally important characterization of a physical
system. In amorphous, and liquid, metals the disorder is by far not
extreme. Instead, neutron scattering data reveal distinct short ranged
order (for a review, see \cite{Waseda}) and the ionic structure factor
displays a series of peaks. Nevertheless, those metals
exhibit a whole range of anomalous properties; especially transport
properties exhibit peculiar behavior, and even the electronic
structure poses a challenge to the theoretician. How to explain the
minimum in the Density of States in noble metal-polyvalent metal
glasses \cite{Haeussler} or even the true gap in the spectrum of
liquid Bi, Tl, Hg \cite{Oelhafen}. What is the
influence of the electronic structure on the transport? On the experimental 
side, detailed analysis of noble metal-polyvalent simple element glasses
\cite{PH} have revealed a strong influence of both the ionic structure on
the electronic one, and of the electronic structure on the ionic one. Namely
when $K_p$ (the location of the first peak of the structure factor) is about $2
k_F$ (the diameter of the Fermi-sphere), a minimum develops in the 
electronic density  
of states, which is best seen in these particular alloys. In turn it causes
the Friedel oscillations of the effective electron-electron interaction
to match the ionic potential (at least as far as the first few minima are 
concerned), thus tending to stabilize the amorphous structure \cite{Beck}
in a manner similar to the crystalline Hume-Rothery phases\cite{Hume,Mott}. 
Recently Kroha {\it et al} 
\cite{Hans} obtained an instability of the Fermi see towards
density oscillations which causes a phase shift of the Friedel
oscillations, as well as an enhancement of their amplitude. This
instability
occurs when the transport time is substantially smaller than the
quasi-particle life-time and is consistent with
experiment. Experimentally this  
behavior is best seen when the valence of the glass $Z$ is close to $1.8$ 
el/Atom.
In these glasses transport coefficients have been systematically investigated
too. As a function of valence, the electrical resistivity is maximum for
$Z \sim 1.8$. There it decreases with temperature, while the thermopower 
grows with the temperature at low temperature. The correlation minimum in
the density of states, negative temperature coefficient of the resistivity
and positive thermopower has not yet found a definite theoretical explanation.

To tackle this problem we apply the Muffin-Tin Effective
Medium Approximation (MT-EMA) \cite{Roth,RothSingh,IBF,IFB}, 
to an
ensemble of atoms, distributed according to the hard sphere structure
factor. Having the electrons moving in the muffin-tin potential
created by the atoms is an appealing description of those all-in-all
rather good metals. The MT-EMA is a self-consistent scheme which is
resumming the disorder averaged multiple scattering series and
addresses the 1-particle and 2-particles quantities on an equal
footing, so as to fulfill the Ward identities. It precisely allows for
a detailed discussion of the interplay of multiple scattering and
short ranged order effects. A first study exists \cite{FBI}. Here we
extend it to a complete discussion of the temperature dependence of
the electrical resistivity and of the thermopower.

\section{ Summary of the formulation}
In this work we investigate the effect of short ranged order (SRO) in
an otherwise topologically disordered system, on the electronic
structure and transport properties of the medium. Such a question is
relevant to metallic glasses and liquid metals, where the periodic
arrangement of the ions is replaced by a structure characterized by
SRO only. In this case a series of properties of the metals are
anomalous. Most striking is the behavior of the electrical
conductivity as a function of the temperature. A compilation of
experimental data by Mooij \cite{Mooij,Howson} leads to the ''Mooij rule``
and separates the alloys into 2 classes of behaviors: first if
the low-temperature resistivity ($\rho_0$) exceeds a certain threshold
value  $\rho_c$ (about 150$\mu\Omega cm$), then increasing the
temperature leads to a {\it decrease} of the resistivity, and second if
$\rho_0<\rho_c$ the more common behavior of seeing the resistivity 
increasing upon an increase of the temperature is
restored. In the meantime, a series of exceptions to the second class
of behavior has been discovered \cite{Cote}, and only the first
behavior holds for good. Quite remarkably $\rho_c$ is not particularly
large, not even 2 orders of magnitude larger than in good metals such
as Cu at room temperature. 
It nevertheless corresponds to a strong reduction of the mean
free path down to several atomic spacings. In this regime weak scattering
approaches are thus expected to be inappropriate.

Another interesting quantity is the thermopower $S$. While at low
temperature a simple estimate using Boltzmann equation yields $S/T =
-{k_b}^2 {\pi}^2/|e| E_F$, experimental evidence points towards a
very different 
situation. Not only the magnitude, but even the sign of $S/T$ are not
universal. They strongly depend on the density of charge carriers.
The Hall conductivity and the magnetoresistance are anomalous, too,
but their investigation is left for future work.

Given the short mean free path, the elucidation of these anomalies is
not expected to follow from a simple weak scattering approach, such as
Ziman formula for the resistivity \cite{Ziman}. Indeed, as noted by
J\"ackle, \cite{Jaeckle}, the amplitude of $S/T$ can be too large to be
obtained out of the Ziman formula, even though the general features
can be understood in this framework. Moreover the experimental finding
of a structure induced minimum in the density of states of noble
metal- polyvalent metal glasses by H\"aussler {\it et al} forces us
to abandon the nearly free electron model. Instead we resort to
the MT-EMA. In this framework free electrons are scattered by the
potential of the ions, and the self-energy (which allows for
interpreting the results in terms of an effective dispersion relation)
is determined by a self-consistent scheme which takes SRO effects into
account, by resumming approximately the disorder averaged multiple
scattering series. This scheme becomes exact in the limit of large
coordination number, provided the Kirkwood superposition principle
holds \cite{Kirkwood}. In real systems the coordination is fairly
large (about 11), and allows us to conclude that we have quite a
reliable approximation. In turn transport coefficients are determined
self-consistently, such as to fulfill Ward identities.

We restrict ourselves to the case of s-scattering only. 
In sum, we first determine the (disorder averaged) Green's
Function:
\begin{equation} 
<G_k(E)> =  G_0(k) + G_{0k}(E) Q_{k}(k,k) G_{0k}(E)
\end{equation}
where the off-shell part of the (disorder averaged) T-matrix is 
related to the on-shell
part at $\kappa^2=E$ by:
\begin{eqnarray}
Q_k(p,p') &=& n
\frac{t(p,p')\tau - t(p,\kappa)t(\kappa,p')}{\tau} \nonumber \\
&+& \frac{t(p,\kappa)}{\tau}Q_k(\kappa,\kappa) \frac{t(p',\kappa)}{\tau}
\end{eqnarray}
the on-shell part of the T-matrix is obtained in terms of its diagonal
part $t_d$ and an effective propagator $\tilde{G}_k(\kappa)$ as:
\begin{equation} 
Q_k(\kappa,\kappa) = \frac{ n t_d(\kappa,\kappa)}{1- n t_d(\kappa,\kappa) 
\tilde{G}_k(\kappa)} \quad .
\end{equation}
The latter 2 quantities are obtained as the solutions of
\begin{eqnarray} \label{on1}
\tilde{G}_k(\kappa) &=& \tilde{B}_k(\kappa) + \int \frac{q^2 dq}{2 \pi^2 n}
a_0(k,q) \tilde{G}_q(\kappa) Q_q(\kappa,\kappa) \tilde{G}_q(\kappa) 
\nonumber \\
t_d(\kappa,\kappa) &=& t(\kappa,\kappa)\nonumber \\ 
+&& \hspace{-0.25cm} t(\kappa,\kappa) 
\int \frac{k^2 dk}{2 \pi^2} \tilde{G}_k(\kappa) Q_k(\kappa,\kappa) B_k(\kappa)
t_d(\kappa,\kappa)
 \end{eqnarray}
where we have introduced $t(p,p')$ as the ion off-shell t-matrix,
$\tilde{B}_k(\kappa)$ the effective propagator in the
Quasi-Crystalline Approximation
\begin{equation}
\tilde{B}_k(\kappa) = B_k(\kappa) + \int \frac{q^2 dq}{2 \pi^2 n} a_0(k,q)
B_q(\kappa)
\end{equation}
with $B_k(\kappa)=1/(\kappa^2-k^2)$ being the free electron propagator.
The information about the structure
enters via the pair distribution function $h(k)$ (which is related to
the structure factor by $S(k) = 1+h(k)$) and the density of the ions
$n$. Expending the pair distribution function according to:
\begin{equation} 
h(|{\bf k}-{{\bf q}}|) = 4 \pi \sum_L Y_L(\hat{k}) a_l(k,q) Y_L(\hat{q})
\end{equation}
defines the quantity $a_0(k,q)$.
We obtain the density of states in the usual way:
\begin{equation}
N(E) = \frac{-2}{n \pi } \int \frac{d{\bf k} }{(2 \pi)^3} \IM(G_k(E))
\end{equation}
and the electronic density by
\begin{equation}
n_e = \int dE N(E) f_F(E-\mu) \quad.
\end{equation}
Next we solve the equations for the transport. The vertex corrections
are determined by solving:
\begin{eqnarray}
X_1(k) &=& \int \frac{q^2 dq}{2 \pi^2 n} a_1(k,q) 
\bigg(
 \Big( \left| \frac{Q_q(\kappa,\kappa)}{n T_d} \right|^2 -1 \Big)
X_1(q)  \nonumber \\ 
&+& \left|\frac{t(q,\kappa)}{t(\kappa,\kappa)} \frac{G_{0q}(E) 
Q_q(\kappa,\kappa)} 
{n T_d} \right|^2 q
\bigg) \nonumber \\
X_2(k,k') &=& \int \frac{q^2 dq}{2 \pi^2 n} a_1(k,q) 
\bigg( \tilde{G}_q(\kappa) Q_q(\kappa,\kappa) X_2(q,k') \nonumber \\
&+&  
\frac{t(q,\kappa)}{t(\kappa,\kappa)} 
\frac{G_{0q}(E) Q_q(\kappa,\kappa)}{n T_d} q G_{0q}(E') \nonumber \\
& & \hspace{1cm} \cdot \frac{ t'(q,k) \tau'
- t'(q,\kappa) t'(\kappa,k)}{t'(\kappa,\kappa)}
\bigg)  \nonumber 
\end{eqnarray}
\begin{eqnarray}
\label{x1}
X_3(k) &=& \int \frac{q^2 dq}{2 \pi^2 n} a_1(k,q) \nonumber \\ & &\cdot
\left| \frac{ t(q,k) 
t(\kappa,\kappa)  
- t(q,\kappa) t(\kappa,k)}{t(\kappa,\kappa)} G_{0q}(E)\right|^2 q 
\end{eqnarray}
out of which we construct:
\begin{eqnarray} 
\Lambda_k(E) &=& 
\left| \frac{t(k,\kappa)}{t(\kappa,\kappa)}Q_{{\bf k}} \right|^2
{X}_1({{\bf k}}) \nonumber \\
&+& 2 \RE \left(\frac{t(k,\kappa)}{t(\kappa,\kappa)} Q_{{k}}
{X}_2({k},k)  \right) 
+ {X}_3({{k}}) 
\end{eqnarray}
which allows for obtaining the $E-$dependent conductivity
\begin{eqnarray} \label{emleit}
\sigma(E) &=& \frac{\hbar}{3 \pi} 
(\frac{e\hbar}{m})^2 \int \frac{k^3 dk}{2 \pi^2}
\bigg( G_{0k}(E^+) G_{0k}(E^-) \Lambda_k(E)  \nonumber \\ 
&-& \frac{m}{\hbar^2} 
\RE \Big( G^2_k(E^+)   
\frac{d\Sigma_k(E^+)}{dk} \Big) \bigg) \nonumber \\
&+& \frac{\hbar}{3 \pi} (\frac{e\hbar}{m})^2 
\int dk k^4 \left(\frac{\IM(G_k(E))}{\pi}\right)^2 
\end{eqnarray}
out of which the conductivity results into:
\begin{equation}
\sigma= \int_{-\infty}^{\infty} d E 
\frac{ - \partial f_F}{\partial E} \sigma(E)
\end{equation}
We shall compare the EMA conductivities to the ones following from
Ziman formula
\begin{equation} 
\rho_{Ziman} = \frac{3\pi n}{4 \hbar e^2 v_F^2 k_F^4} \int_0^{2 k_F}
dq q^3 S(q) |t(\kappa,\kappa)(E_F)|^2 \quad.
\end{equation}
In linear response the thermopower S results into:
\begin{equation} \label{thp}
S =-\frac{1}{eT} \frac{
\int_{-\infty}^{\infty} d E 
\frac{ - \partial f_F}{\partial E} (E-\mu)
\sigma(E)}{\int_{-\infty}^{\infty} d E  
\frac{ - \partial f_F}{\partial E} \sigma(E)}
\end{equation}
For more details we refer to \cite{IBF,dipuli}.

\section{Results}

\begin{figure}
\narrowtext
\centerline{\epsfxsize=8.5cm \epsfbox{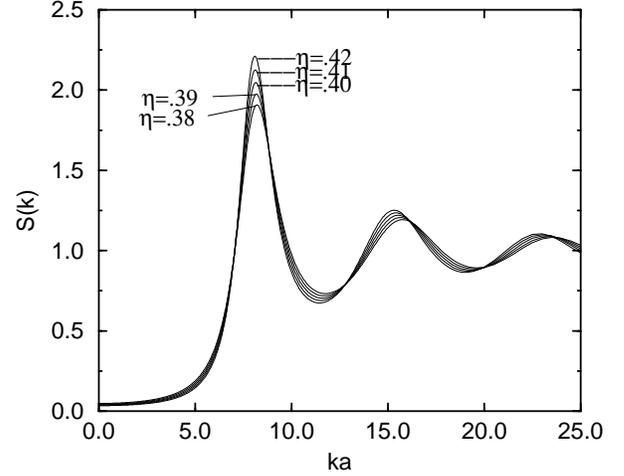}}
\caption{\label{fig1} The hard sphere structure factor for different
packing fractions.} 
\end{figure}

\begin{figure}
\narrowtext
\centerline{\epsfxsize=8.5cm \epsfbox{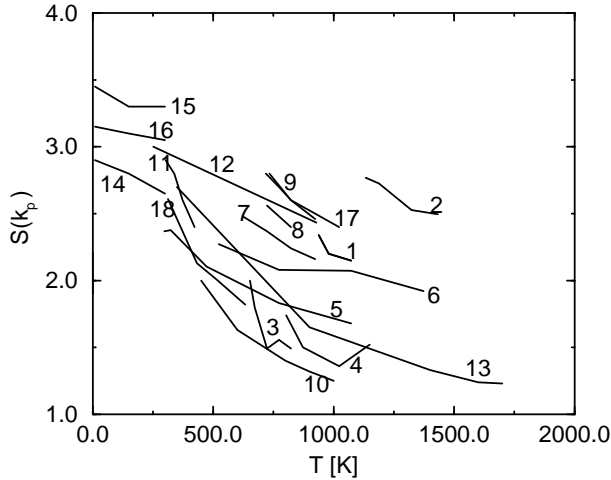}}
\caption{\label{fig2} The amplitude of the first peak of the structure
factor $S(k_p)$ as  
function of temperature. 
[29]
1: liq. Al (6), 
2: liq. CuSn (1), 
3: liq. GeTe (2), 
4: liq. AlGe (3), 
5: liq. Ga (4), 
6: liq. Sn (4), 
7: liq. Cd (4), 
8: liq. Zn (4), 
9: liq. Zn (5), 
10: liq. Rb (7), 
11: liq. Cs (8), 
12: liq. Ga (9), 
13: liq. Rb (10), 
14: amor. MgZn (11), 
15: amor. CaZn (11), 
16: amor. MgCu (11), 
17: liq. Zn (12), 
18: liq. Rb [23].
}
\end{figure}

We first describe how the atomic structure is taken into account. Here
we resort to the Percus-Yevick solution of the hard sphere liquid to
model the ionic structure. The resulting structure factor is displayed
in fig.~\ref{fig1}, for packing fractions $\eta$ ranging from $\eta = .38$ to
$\eta = .42$, which are reasonable values in order to account for at
least liquid simple metals. 
Even though merely a model structure
factor, it is expected to capture the main features of the structure,
and has been favorably compared to the one of liquid simple metals
\cite{AshcroftL}. It also has the advantage of substantially reducing
the numerical effort since $h(k)$ has an analytical expression. Here
the height of the first peak of the structure factor can be varied by
tuning the packing fraction, in order to mimic a change in
temperature. In order to model the dependence of $\eta$ on $T$, we
have reported the height of  the first peak of the structure factor
$S(k_p)$ as a function of temperature for various systems \cite{stru1}in
fig.~\ref{fig2}. Applying a best fit procedure reveals that $S_T(k_p)$ obeys a
law:
\begin{equation}
S_T(k_p) = \frac{1}{aT+b}
\end{equation}
sticking to liquid Rb for the sake of definiteness yields a=5.09
$10^{-4} K^{-1}$ and $b=0.23378$. We can then model the temperature
dependence of $\eta(T)$, listed in table 1.
\begin{table}
\caption{Temperature dependence of the packing fraction}
\begin{tabular}{|l|c|c|c|c|c|}  \hline
$S(k_p)$& 2.208 & 2.129& 2.049& 1.970 & 1.901 \\ \hline
$\eta$ &.42&.41&.40&.39&.38 \\ \hline
$T[K]$ &430 &463&499&537 & 573  \\ \hline
\end{tabular} 
\end{table}
We use the square well potential to model the screened Coulomb
potential of the ions. In atomic units 
it is described by the parameter $V_z=V a^2$,
where $V$ is the depth of the potential and $2a$ its range. Throughout this
work we use $V_z=1.4$ and the unit $\hbar=2m=e^2=1$. 

\subsection{Electronic structure}

\begin{figure}
\narrowtext
\centerline{\epsfxsize=8.5cm \epsfbox{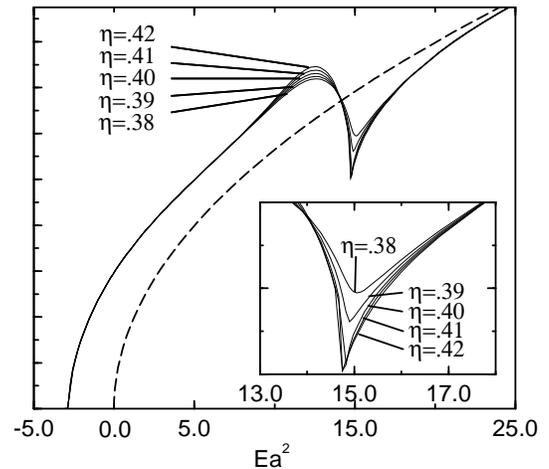}}
\caption{\label{fig3} The electronic density of states, for the
potential strength  
$V_z=1.4$ and for different packing fractions. The dashed curve corresponds 
to the calculated free electron density of states. Inset: expanded view around
$E_m$.}
\end{figure}

We now proceed to solve the set of equations (\ref{on1}), determining
the electronic structure. To some extent these equations have been
solved by Fr\'esard {\it et al} \cite{FBI}. Here we solve them in order to first
determine the parameter range where the theory is analytic. Indeed the
breakdown of analyticity of the MT-EMA has been observed for the
delta-function potential by Singh and Roth \cite{SinghRoth} possibly yielding
negative spectral functions especially for large packing fractions, and no
proof of analyticity of the approximation has been found \cite{Roth2,Itoh}.
It turns out that the analyticity breaks down for low Fermi
energies, when $\eta>0.44$, namely a value which is above those
we are dealing with in this work. For $\eta\leq0.42$, the theory is
analytic for all values of the potential and of the Fermi energy. The
main result concerns the density of states and is displayed in
fig.~\ref{fig3}. 2 features appear 
clearly.\\ 

First a deep minimum, and second quite a sizeable peak just
below it. The minimum is becoming deeper and deeper as $\eta$ is
growing and occurs at $E_m$, with $E_m \sim (k_p/2)^2 $ with little
dependence on the potential strength. It is thus fair to say that it
results from quasi Bragg scattering on a quasi zone boundary. It
replaces the energy gap which opens when the lattice periodicity is
restored. As a result breaking the periodicity of the lattice does not
invalidate the concept of zone boundary as a whole: even though the
latter cannot be defined as the inverse of the lattice spacing but is
rather given in an effective way by about $k_p/2$, it still locates
the region in k-space where a pseudo-gap opens due to strong
back-scattering. 
At $E_m$, and below it, the effective
dispersion flattens which leads to the maximum in the density of
states. Such a maximum leads to a decrease in the total energy 
making it very likely to find most real liquid and amorphous metals
with a Fermi energy close to $E_m$, in agreement with the argument of Nagel 
and Tauc \cite{Nagel}. 
\begin{figure}
\narrowtext
\centerline{\epsfxsize=8.5cm \epsfbox{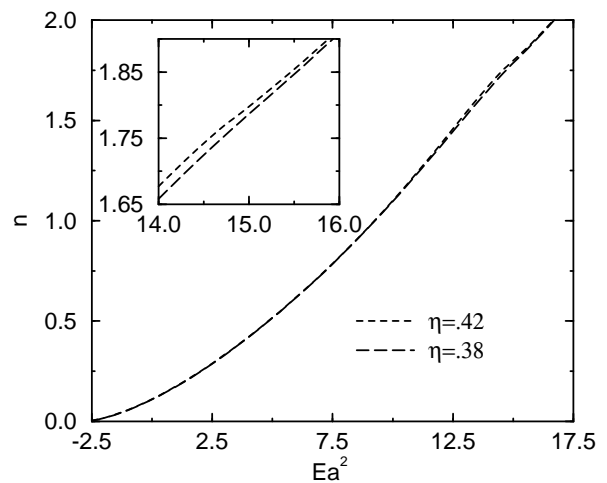}}
\nopagebreak
\caption{ \label{fig4} The density of electrons, for the potential strength 
$V_z=1.4$ and for different packing fractions. Inset: expanded view around
$E_m$.}
\end{figure}
The dependence of the charge
carrier density on the Fermi energy is displayed in fig.~\ref{fig4} , and it
appears that $E_m$ corresponds to $n_m=1.78$.
This value is not
universal, and depends on both $V_z$ (clearly) and $\eta$ (weakly).
Typically it will be shifted down to lower densities by reducing
$V_z$, or increasing $\eta$. Note that the lower band edge is shifted
down from 0 due to the averaged attractive potential. More details
concerning the spectral function, the self-energy and the effective
dispersion can be found in \cite{FBI,dipuli}.

\subsection{Electrical Resistivity}

\begin{figure} 
\narrowtext
\centerline{\epsfxsize=8.5cm \epsfbox{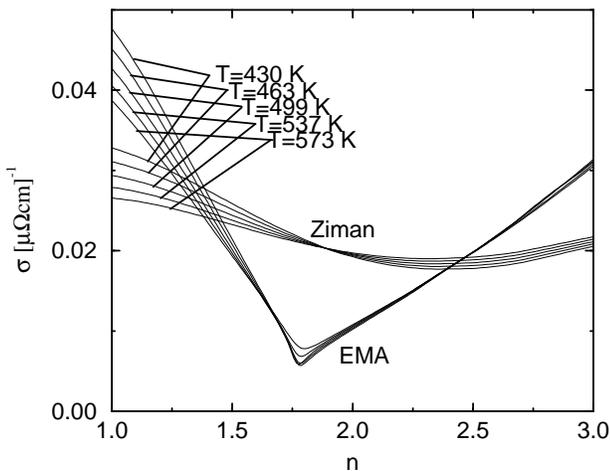}}
\caption{\label{fig5} The electrical conductivity, for the potential 
strength $V_z=1.4$ at different temperatures, calculated in
EMA and with the Ziman formula.}
\end{figure}

We now proceed to the solution of the transport equations Eq.
(\ref{x1}). The result is displayed in fig.~\ref{fig5}, where we display the
conductivity as a function of the electronic density for several
temperatures. 
For each temperature the conductivity has a deep minimum
at $n_e \sim 1.8$, corresponding to $E_m$ (see above). This minimum does
not only follow from the short life time of the quasi-particles, but
from large vertex corrections too. It becomes deeper and deeper as the
temperature is lowered. Note that the bundle of curves crosses at $n_e
\sim 1.5$ and $n_e \sim 2.5$, namely where the bundle of density of
states crosses too. 

This is the root of the Mooij rule. Due to strong multiple scattering
at the pseudo zone boundary, a pseudo-gap opens in the 1-particle
spectrum, which becomes deeper and deeper as SRO grows. In turn it
induces a minimum in the conductivity which itself is getting deeper
and deeper as SRO grows, namely as the temperature decreases. It thus
turns out that the resistivity decreases as the temperature increases,
when the resistivity is large. Contrary to that, left from the
minimum, the effect of the pseudogap results into the flattening of
the effective dispersion, leading to the maximum in the density of
states, which becomes higher and higher as SRO grows, and so does
the conductivity. It works however in the opposite direction right
from the minimum, and as a result the resistivity can both increase or
decrease as the temperature increases,
when the resistivity is small.\\
The contributions to $\rho$ (see eq. (\ref{emleit})) are consisting
of a contribution from the 
spectral function and a series of vertex corrections. It turns out
that the latter are very importent, and can increase the contribution
from the spectral function by as much as a factor $2$. For a more
detailed discussion see \cite{FBI,dipuli}.

\begin{figure}
\narrowtext
\centerline{\hspace{1cm} \epsfxsize=8.5cm \epsfbox{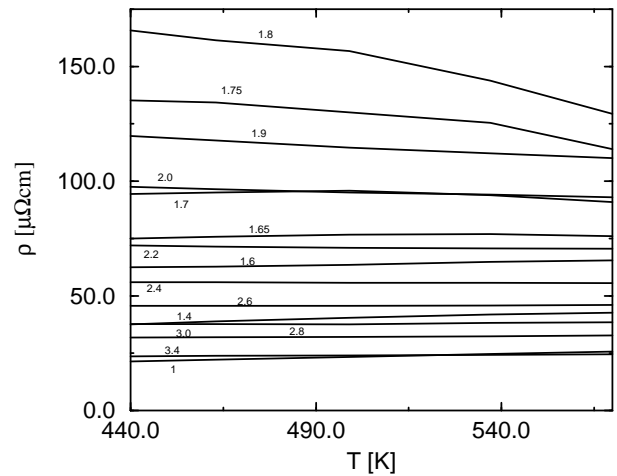}}
\caption{ \label{fig6} The resistivity as function of temperature, for the 
potential strength $V_z=1.4$ and for different
densities of electrons.}
\end{figure}

\begin{figure} 
\narrowtext
\centerline{\hspace{1cm} \epsfxsize=8.5cm \epsfbox{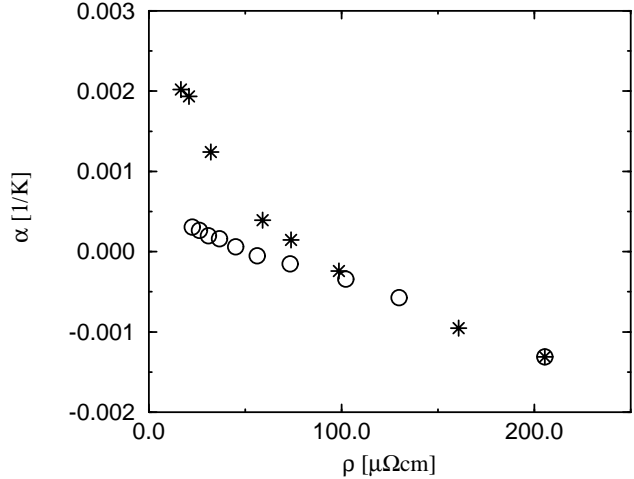}}
\caption{\label{fig7} The Mooij--correlation for the potential 
strength $V_z=1.4$ at $300 K$, 
$(\circ)$ corresponds to $E_F<E_m$
and 
$(\star)$ to $E_F>E_m$. The minimum of $\alpha$ is correlated to 
$E_F\sim E_m$ at $n_e=1.78$.
}
\end{figure}

In more detail the temperature dependence of the resistivity is
displayed in fig.~\ref{fig6} for several values of the electronic
density.  In the regime of low electronic density $(n_e \leq 1.7)$
where $E_F<E_m$, the resistivity 
increases with temperature at a moderate rate. When the electronic
density $n_e$ approaches
$n_m$ than the resistivity $\rho$ decreases with increasing
temperature $T$ at a higher rate. It is maximum
at $n_m$. A further increase in the density leads to a decrease of this rate,
for $n_e$ up to $2.4$ where it changes its sign. 
Those data can be summarized by reporting the 
temperature coefficient of the resistivity $\alpha$:
\begin{equation}
\alpha \equiv \frac{1}{\rho} \frac{\partial \rho}{\partial T}
\end{equation}
as a function of the resistivity. 
Calculating $\alpha$ at the lowest temperature leads to fig.~\ref{fig7} . 
The data scatter onto 2 lines which represent the
2 regimes where the Fermi energy is either smaller or larger than
$E_m$. Not only the sign of $\alpha$, but its amplitude as well,
compare favorably with the experiment \cite{Mooij,Howson}. In contrast 
to ref. \cite{FBI}
we have calculated the temperature dependence of the resistivity for
fixed density. There are 3 contributions to $\alpha$: (i) the
$\eta$-dependence of the resistivity as such, (ii) the smearing of the
Fermi surface together with a strong $E$-dependence of
$\sigma(E)$, (iii) the change in the Fermi energy (at fixed density)
following from a change in $\eta$ and $T$.

The first contribution to $\alpha$ is dominated by 1-particle effects
and has been discussed above. This is the leading one.
The second one results from the
curvature of $\sigma(E)$ as one can infer it from the Sommerfeld
expansion, and is contributing to $\alpha$ in much the same way as the
first one; namely the curvature of $\sigma(E)$ is positive in the
vicinity of $E_m$, negative below it, and very small above it. 
This is the second leading one. The
third contribution, however, works in the opposite direction, and
reduces the value of $\alpha$ at $E_m$.

\subsection{Thermopower}

\begin{figure} 
\narrowtext
\centerline{\hspace{1cm}\epsfxsize=8.5cm \epsfbox{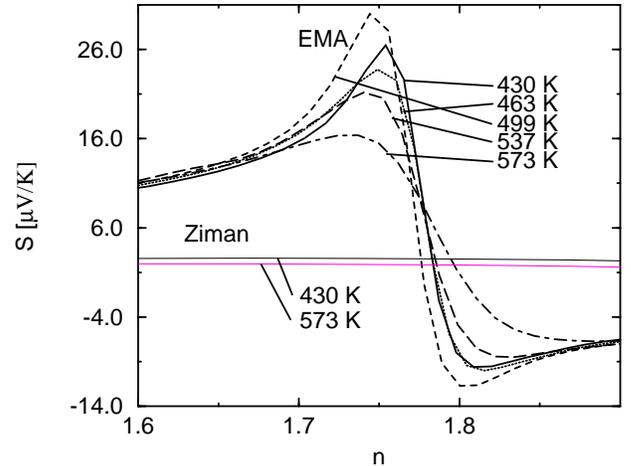}}
\caption{\label{fig8} The thermopower, for the potential strength $V_z=1.4$ 
and at different temperatures, calculated in
EMA and with the Ziman formula.}
\end{figure}

We now evaluate the thermopower according to Eq. (\ref{thp}) and we
display it in fig.~\ref{fig8} where it is compared to the results obtained out
of the Ziman formula. 

As
compared to the results obtained out of the Ziman formula, we obtain much
larger values, either positive if $E_F<E_m$ or negative if $E_F>E_m$.
Again the position of the Fermi energy with respect to $E_m$ plays the
key role for assessing the sign of the thermopower. It changes its
sign when the resistivity reaches its maximum value. 
Thus phononic
mechanisms are not required to explain the experimentally observed
large values of $S$, even though they are certainly going to play an
important role as well, especially around the Debye
temperature. Another important contribution, namely the phonon drag
effect, is however small in metallic glasses \cite{Jaeckle}, because the 
scattering of the phonons by the high degree of disorder keeps them essentially
in local thermal equilibrium.

\begin{figure} 
\narrowtext
\centerline{\hspace{1cm}\epsfxsize=8.5cm \epsfbox{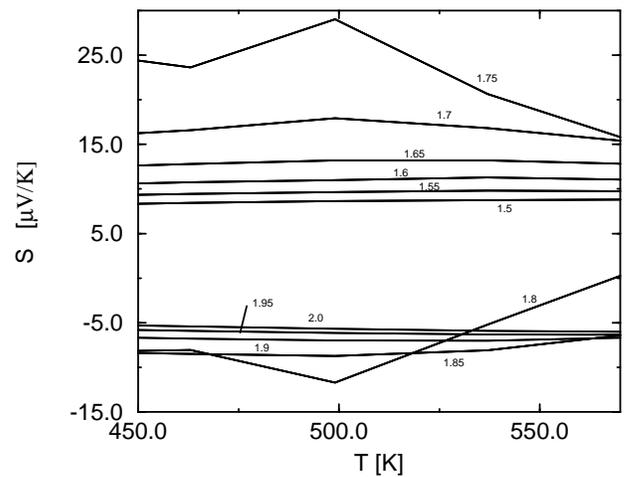}}
\caption{\label{fig9} The thermopower as function of temperature, for
the potential  
strength $V_z=1.4$ and for different
densities of electrons.}
\end{figure}

We turn to the $T$-dependence of $S$ and display it on fig.~\ref{fig9}.
 Typically it is growing with the
temperature when the Fermi energy is smaller than $E_m$, whereas it is
decreasing when $E_F>E_m$. If $E_F$ is within $k_BT$ of $E_m$, there
is a crossover region where $\partial S/\partial T$ can have either
sign. Namely, if $E_F<E_m$, we have $\partial S/\partial T > 0$ if $k_BT
< E_F-E_m$ and  
$\partial S/\partial T < 0$ if $k_BT > E_F-E_m$. On the other hand, if
$E_F<E_m$, we have 
$\partial S/\partial T < 0$ if $k_BT < E_F-E_m$ and 
$\partial S/\partial T > 0$ if $k_BT > E_F-E_m$. Such a
behavior has been observed in noble metals- polyvalent metals alloys such as
(Au, Ag, Cu) - Sn. Indeed increasing the content of Sn (i.e. the density of
charge carriers) changes the sign of the thermopower at low temperature from
positive to negative while the resistivity decreases as review by H\"aussler
\cite{Haeusslerpr}. We finally note that the values of S are too large to fit
the experiments. This follows from the relatively sharp E-dependence of the
density of states, which is strongly reflected in the thermopower. The 
situation should improve by including multiple occupancy corrections to the
theory, but this is beyond the scope of this work.
\par
%
\section{Summary and acknowledgment}

In summary we have solved the MT-EMA equations for a system of
s-scatterers. The results lead us to a consistent interpretation of
the anomalous transport properties of amorphous metals. Especially we
found that if the resistivity exceeds a certain threshold, than it
decreases with increasing temperature, while otherwise the temperature
coefficient of the resistivity can have either signs. 
This behavior follows from the
opening of a pseudo-gap at the quasi zone boundary. The pseudo-gap is
responsible for a large resistivity, as well as for a large
thermopower (which can be 
positive or negative) and its non-trivial temperature dependence. As a
function of the charge carrier density, the low temperature
thermopower changes its
sign when the resistivity reaches its maximum.

It is a pleasure to acknowledge Prof.~P.~W\"olfle, Prof. H. Beck 
for enlightening discussions, 
and Prof.~M.~Itoh for useful discussions.
RF gratefully
acknowledges financial support from the Deutsche
Forschungsgemeinschaft
under SFB 195, and the Japanese Science Foundation under grant A 07740301.

\section{Appendix}
In order to solve the set of equations Eq.(\ref{on1}) numerically we
use a set of transformations in order to make the task
tractable. First we solve the equations for complex energies with
$\IM(E)=1$ which turns out to be easy to solve. We then gradually
reduce $\IM(E)$ down to $10^{-4}$, a value that is typically smaller
than $-\IM(\Sigma(k_F))$ which has thus very little influence on the
result. The equations will be solved by iteration. Even though we use a
linear combination of the previous 3 iterations to improve the
convergence, the combination being chosen so as to optimize the
difference between the input and the output, the task turn out to be
very tedious and require as much as $ 10^3$ iterations for each
value of $\eta$ and $E_F$. We handle the on-shell singularity which
enters the expression for $\tilde{G}_k(\kappa)$ in the same way as
described in ref. \cite{FBI}, namely we introduce:
\begin{eqnarray} 
\widehat{\tilde{G}}_k(\kappa) &\equiv& \tilde{G}_k(\kappa) - B_k(\kappa)
\nonumber \\
\Sigma_d &\equiv& \frac{1}{t(\kappa,\kappa)} - \frac{1}{t_d(\kappa,\kappa)}
\end{eqnarray}
which obeys the equations:        
\begin{eqnarray} 
\widehat{\tilde{G}}_k(\kappa) &=& \int \frac{q^2 dq}{2 \pi^2 n}
a_0(k,q)
\nonumber \\ & & \hspace{-1.35cm} \cdot     
\frac{1-t(\kappa,\kappa) \Sigma_d + n t(\kappa,\kappa) B^{-1}_q(\kappa)
\widehat{\tilde{G}}_q(\kappa)^2 
+ n t(\kappa,\kappa) \widehat{\tilde{G}}_q(\kappa) }
{B^{-1}_q(\kappa) (1-t(\kappa,\kappa)
(\Sigma_d+n\widehat{\tilde{G}}_q(\kappa))) -n t(\kappa,\kappa)} 
\nonumber \end{eqnarray}
\begin{eqnarray} \hspace{-1.5cm}
\Sigma_d &=& n t(\kappa,\kappa) \int 
\frac{k^2 dk}{2 \pi^2} \nonumber \\
& & \hspace{-0.5cm} \cdot 
\frac{ B_k(\kappa) + \widehat{\tilde{G}}_k(\kappa) }
{B^{-1}_k(\kappa) (1-t(\kappa,\kappa)
(\Sigma_d+n\widehat{\tilde{G}}_k(\kappa))) -n t(\kappa,\kappa)} 
\end{eqnarray}

Nevertheless, even though the new
equation is now smooth on-shell, it will develop a singular behavior
at the renormalized quasi-particle peak through the T-matrix. The
difference is that the location of the pole is not known beforehand,
and lies in the complex plane. Thus some additional manipulations of
the equations is required. They are inspired by considering the
integral:
\begin{equation} \label{ap1}
\int \frac{f(x)}{x-z} dx \quad. 
\end{equation} 
The integrand is singular at z, z being complex. If $\IM(z)$ is large,
the resulting integrand is smooth and such an integral can be
evaluated as such. But when $\IM(z)$ is small, it is best to transform
the integral Eq. (\ref{ap1})
\begin{equation}
\int \frac{f(x)}{x-z} dx = \int \frac{f(x)-f(z)}{x-z} dx + 
f(z)\int \frac{1} {x-z} dx 
\end{equation} 
and evaluate the second term analytically. Obviously the resulting
integrand is smooth. Applying this principle to Eq.(\ref{on1}) allowed
us to greatly improve the numerical stability of the scheme.

Another difficulty arises because the integrands in Eq.(\ref{on1})
decrease slowly for large momenta. In this case we have:
\begin{eqnarray}
&&\frac{1-t(\kappa,\kappa) \Sigma_d + n t(\kappa,\kappa) B^{-1}_q(\kappa)
\widehat{\tilde{G}}_q(\kappa)^2 
+ n t(\kappa,\kappa) \widehat{\tilde{G}}_q(\kappa) }
{B^{-1}_q(\kappa) (1-t(\kappa,\kappa)
(\Sigma_d+n\widehat{\tilde{G}}_q(\kappa))) 
-n t(\kappa,\kappa)} \nonumber \\ & & \cdot
\frac{q^2 dq}{2 \pi^2 n} a_0(k,q)    
 \stackrel{q\gg k}{\longrightarrow}
\frac{-1}{2 \pi^2 n} \frac{a_0(k,q)}{q^2} 
\quad.
\end{eqnarray}
Defining
\begin{equation} \label{defhhk}
H(k) \equiv \int_{k_{cutoff}}^{\infty} \frac{-dq}{2 \pi^2 n} 
\frac{a_0(k,q)}{q^2} 
\end{equation}
we also obtain the large momenta contribution to $\Sigma_d$ as
\begin{equation} \label{deftheta}
\theta_d \equiv \frac{-n t(\kappa,\kappa)}{t(\kappa,\kappa) \Sigma_d
-1}
\int_{k_{cutoff}}^{\infty} 
\frac{dk}{2 \pi^2} (B_k(\kappa) + H(k))
\end{equation}
which is added to $\Sigma_d$ at each iteration, in the same way as
$H(k)$ is added to $\widehat{\tilde{G}}_k(\kappa)$. Both integrals
entering Eq. (\ref{defhhk}, \ref{deftheta}) can be calculated once and for all.

\end{multicols}


\begin{thebibliography}{52}
\bibitem[*]{bylines} on leave of absence from, Physics Department, Shimane
University, Nishikawatsu-cho 1060, Matsue 690, Shimane, Japan
\bibitem[+]{bylines} Work partially performed at: Institut f{\"u}r Theorie 
der Kondensierten Materie, Physikhochhaus, Universit{\"a}t Karls\-ruhe,
Postfach 6980, D-76128 Karlsruhe
\bibitem{aral} E.~Abrahams, P.W.~Anderson, D.C.~Liciardello and 
T.V.~Ramakrishnan, Phys. Rev. Lett. {\bf 42}, 673 (1979).
\bibitem{VW} D.~Vollhardt, P.~W\"olfle, {\sl Self-Consistent Theory of
Anderson Localization}, Electronic Phase Transitions, ed.~W.~Hanke, 
Y.~V.~Kopaev,  ESP, (1992).
\bibitem{Waseda} J. F.~Sadoc, C.N.J.~Wagner, Glassy
Metals II, {\bf 51}, ed. H.~Beck, H.-J.~G\"untherodt, Springer 
(1983).
\bibitem{Haeussler}P.~H\"aussler {\it et al} , 
Phys. Rev. Let., {\bf 51}, No 8, 714 (1983); P.~H\"aussler {\it et al}, 
Journal of Non-Cryst. Solids {\bf 61 \& 62}, 1249 (1984); 
P.~H\"aussler, {\it et al}, Europhys. Lett. {\bf 15}, 759 (1991).
\bibitem{Oelhafen} P. Oelhafen, G. Indlekofer and H.-J.~G\"untherodt, 
Proceedings of the LAM 6
conference, Garmisch-Partenkirchen (1986), ed. W.~Gl\"aser, F.~Henzel 
and E.~L\"uscher, Z. Phys. Chem. Neue Folge {\bf 157}, 483 (1988).
\bibitem{PH}P.~H\"aussler, Glassy
Metals III, p. 163, ed. H.~Beck, H.-J.~G\"untherodt , Springer 
(1994).
\bibitem{Beck} H.~Beck, R.~Oberle {\it Proc. 3rd Int'l Conf. on Rapidly 
Quenched 
etals}, ed. by B. Cantor (The Metal Society, London 1978) Vol.1, p. 416; 
R.~Oberle, H.~Beck: Solid State Commun. {\bf 32}, 959 (1979); H.~Beck,
R.~Oberle: 
J. Phys. {\bf 41}, C8, 289 (1980).
\bibitem{Hume} W.~Hume-Rothery: {\it The Structure of Metals and Alloys} 
Inst. Metals, monogr. Rep. Ser. No. 1, London, 1956.
\bibitem{Mott} N.F.~Mott, H.~Jones: {\it The Theories of the
Properties of Metals and Alloys}, Clarendon, Oxford 1936, p. 310. 
\bibitem{Hans} J.~Kroha, A.~Huck, T.~Kopp, Phys. Rev. Lett. {\bf 75},
4278 (1995).
\bibitem{Roth}L.~Roth, Phys. Rev. B {\bf 9}, 2476 (1974).
\bibitem{RothSingh}V.~Singh, L.~Roth, Phys. Rev. B {\bf 25}, 2522 (1982).
\bibitem{IBF}M.~Itoh, R.~Fr\'esard, H.~Beck, J.~Phys. Cond. Mat. {\bf 1}, 6381
(1989). 
\bibitem{IFB}M.~Itoh, R.~Fr\'esard, H.~Beck, J. Phys. Cond. Mat. 
{\bf 2}, 2687  (1990).   
\bibitem{FBI} R.~Fr\'esard, H.~Beck, M.~Itoh, J. Phys. Cond. 
Mat. {\bf 2}, 8827 (1990).
\bibitem{Mooij}J.~H.~Mooij, Phys. Stat. Sol. {\bf A17}, 521 (1973). 
\bibitem{Howson} M.A.~Howson and B.L.~Gallagher, Phys. Rep.  {\bf 170} (5), 266
(1988).
\bibitem{Cote}P.~J.~Cote, L.~V.~Meisel, Glassy
Metals I, p. 141, ed. H.~Beck, H.-J.~G\"untherodt, Springer 
(1981); P.~L.~Rossiter, Cambridge University Press (1987);
U.~Mizutani, Progr. in Materials Science {\bf 28}, 97 (1983/1985).
\bibitem{Ziman} J.M.~Ziman, Avd. Phys. {\bf 16}, 551 (1967);
O.~Dreirach, R.~Evans, H.-J.~G\"untherodt, H.~U.~K\"unzi,
J. Phys. F: Metal Phys., {\bf 9}, 505 (1979). 
\bibitem{Jaeckle}J.~J\"ackle,  J. Phys. F: Metal
Phys., {\bf 10}, L43 (1980).
\bibitem{Kirkwood} S.A.~Rice and P.~Gray, {\it The Statistical mechanics
 of Simple Metals (Wiley, New-York, 1965)}, p. 72.
\bibitem{dipuli}U.~Werner, Diploma Thesis, Karlsruhe University (unpublished).
\bibitem{AshcroftL}N.~W.~Ashcroft, J.~Lekner,  Phys. Rev. {\bf 145}, 1 (1966).
\bibitem{SinghRoth}V.~Singh, L.~Roth , Phys. Rev. B {\bf 21}, 4403 (1980).
\bibitem{Roth2}L.~Roth, Phys. Rev. B {\bf 22}, 2793 (1980).
\bibitem{Itoh}M.~Itoh,
unpublished.
\bibitem{Haeusslerpr} P. H\"aussler, Phys. Rep. {\bf 222}, 125 (1992).
\bibitem{Nagel} S.R.~Nagel, J.~Tauc: Phys. Rev. Lett. {\bf 35}, 380 (1975); 
Solid State Commun. {\bf 21}, 129 (1977).
\bibitem{stru1} {\sl Experimental data for the structure factor} \newline 
(1) G.~Busch, H.~J.~G\"untherodt,  Winterkolloquium DPG (1973). \newline
(2) W.~Matz, W.~Meyer, Phys. {\sl Amorphous structures}, ed. D.~Schulze,
Akademie-Verlag (1990). \newline
(3) J.-G.~Gasser, M.~Mayoufi, M.-C.~Belissent-Funel, J. Phys.:
Condens. Matter {\bf 1}, 2409 (1989). \newline
(4) C.~N.~J.~Wagner, Inst. Phys. Conf. Ser. No. {\bf 30}, 110 (1977).
\newline 
(5) W.~Knoll, Inst. Phys. Conf. Ser. No. {\bf 30}, 117 (1977).
\newline
(6) D.~Jovi\'c, I.~Pudureanu , S.~Rupeanu, Inst. Phys. Conf. Ser. No. {\bf
30}, 120 (1977). \newline 
(7) R.~Block, J.-B.~Suck, W.~Freyland, F.~Hensel, W.~Gl\"aser,
Inst. Phys. Conf. Ser. No. {\bf 
30}, 126 (1977). \newline 
(8) M.~J.~Huijben, W.~van der Lugt, Inst. Phys. Conf. Ser. No. {\bf
30}, 142 (1977). \newline 
(9) H.~Beck R.~Oberle, Journ. de Physique, {\bf C8}, 289 (1980). \newline
(10) G.~Franz, W.~Freyland, W.~Gl\"aser, F.~Hensel, E.~Schneider, Journ. de
Physique, {\bf C8}, 194 (1980). \newline 
(11) W.~From, W.~B.~Muir, Phys. Rev. B, {\bf 45}, No 2, 673 (1992). \newline
(12) G.~Etherington, C.~N.~J.~Wagner, Journal of Non-Cryst. Solids, 
{\bf 61 \& 62}, 325 (1984). 

\end{thebibliography}
\end{document}